%% file: Cu3Au-Prep-paper.tex
\def\printcomments{1}

\def\printsubsec{0}

\documentclass[3p, twocolumn]{elsarticle}

\usepackage{amssymb}

\usepackage[T1]{fontenc}
\usepackage[utf8]{inputenc}
\usepackage[retain-unity-mantissa = false,detect-all]{siunitx}
\usepackage[version=4]{mhchem}
\usepackage{hyperref}
\usepackage[dvipsnames]{xcolor}

\usepackage{helvet}

\newcommand{\K}[1]{\SI{#1}{\kelvin}}
\newcommand{\uA}[1]{\SI{#1}{\micro\ampere}}
\newcommand{\nA}[1]{\SI{#1}{\nano\ampere}}
\newcommand{\mbar}[1]{\SI{#1}{\milli\bar}}
\newcommand{\mV}[1]{\SI{#1}{\milli\volt}}
\newcommand{\V}[1]{\SI{#1}{\volt}}
\newcommand{\kV}[1]{\SI{#1}{\kilo\volt}}
\newcommand{\minutes}[1]{\SI{#1}{\minute}}
\newcommand{\seconds}[1]{\SI{#1}{\second}}
\newcommand{\nm}[1]{\SI{#1}{\nano\metre}}
\newcommand{\angdeg}[1]{\SI{#1}{\degree}}

\newcommand{\Lonetwo}{$\text{L}1_2$\ }

\include{figures}

\newcommand{\Sec}[1]{\section{#1}}
\newcommand{\Ssec}[1]{}
\newcommand{\Sssec}[1]{}

\if\printsubsec1
    \renewcommand{\Ssec}[1]{\subsection{#1}}
    \renewcommand{\Sssec}[1]{\subsubsection{#1}}
\fi

\newcommand{\JGC}[1]{}
\newcommand{\DCC}[1]{}
\newcommand{\SOC}[1]{}

\if\printcomments1
    \renewcommand{\JGC}[1]{{\em \noindent \color{red} JG: #1}}
    \renewcommand{\DCC}[1]{{\em \noindent \color{red} DC: #1}}
    \renewcommand{\SOC}[1]{{\em \noindent \color{red} SO: #1}}
\fi

\begin{document}

\begin{frontmatter}




\title{Large insulating nitride islands on Cu$_3$Au as a template for atomic spin structures}


\author{Jeremie Gobeil}
\author{David Coffey}
\author{Shang Jen Wang}
\author{Alexander F. Otte\corref{cor1}}

\ead{a.f.otte@tudelft.nl}
\cortext[cor1]{Corresponding author.}

\address{Department of Quantum Nanoscience, Kavli Institute of Nanoscience, Delft University of Technology, Lorentzweg 1, 2628 CJ Delft, The Netherlands}

\begin{abstract}
We present controlled growth of c(2$\times$2)N islands on the (100) surface of Cu$_3$Au, which can be used as an insulating surface template for manipulation of magnetic adatoms. Compared to the commonly used Cu(100)/c(2$\times$2)N surface, where island sizes do not exceed several nanometers due to strain limitation, the current system provides better lattice matching between metal and adsorption layer, allowing larger unstrained islands to be formed. We show that we can achieve island sizes ranging from tens to hundreds of nanometers, increasing the potential building area by a factor 10$^3$. Initial manipulation attempts show no observable difference in adatom behaviour, either in manipulation or spectroscopy.
\end{abstract}


\end{frontmatter}



\Sec{Introduction}
\label{sec:intro}
The ability to position individual magnetic adatoms into a specific arrangement on a surface holds great potential for atomic scale studies of quantum magnetism \cite{spinelli_imaging_2014}. A particularly successful template for the placement of transition metal atoms is the c(2$\times$2) reconstruction of nitrogen on the \ce{Cu(100)} crystal surface \cite{leibsle_stm_1994}, which provides a self-terminated insulating monolayer, separating the atomic spins from the conduction electrons in the metal below \cite{hirjibehedin_spin_2006}. Due to its covalent structure, the copper-nitride surface provides significant magneto-crystalline anisotropy \cite{hirjibehedin_large_2007} and allows for tunable spin-spin coupling between neighbouring atoms, both ferromagnetic and antiferromagnetic \cite{bryant_local_2013, spinelli_exploring_2015}. The combination of these techniques has given rise to a range of seminal experiments, including the construction of a 96-atom magnetic byte \cite{loth_bistability_2012}, the observation of spin waves in a one-dimensional spin chain \cite{spinelli_atomically_2015}, and the atomically precise study of various highly entangled spin systems \cite{toskovic_atomic_2016, choi_building_2017}.

As atom manipulation techniques become more reliable \cite{kalff_kilobyte_2016}, the size of atomic structures is only limited by the maximum available continuous building area. In the case of copper-nitride, this limit is imposed by the nitrogen islands. Due to a 3\% lattice mismatch between the adsorption layer and the underlying \ce{Cu(100)} crystal, island sizes are strain-limited to \nm{\sim 5}~$\times$~\nm{5} --- or, on saturated surfaces up to \nm{20}~$\times$~\nm{20} \cite{oberg_control_2014-1} --- hampering the assembly of any spin structure larger than that. Here, we present growth of nitride islands on a different metal substrate: the \ce{Cu3Au(100)} surface. With a lattice constant $a=\nm{0.375}$ \cite{niehus_surface_1993}, its lattice much better matches the one of copper-nitride ($a=\nm{0.372}$ \cite{choi_incommensurability_2008}) than the \ce{Cu(100)} surface ($a=\nm{0.359}$ \cite{davey_precision_1925}) does. By properly tuning growth conditions, we can routinely grow islands ranging from tens to hundreds of nanometres across, vastly increasing the area on which spin structures can be assembled.

\figOrdered
\figDisordered

\Sec{Experimental details}
\label{sec:experimetal}

The experiments were performed in a scanning tunnelling microscope (STM) operating in ultra-high vacuum (UHV) and cryogenic conditions. During measurements the pressure was \mbar{<5e-10} and the temperature was between \K{1.4} and \K{1.5}. Sample preparation was performed \emph{in situ} in a UHV chamber connected to the STM, which has a base pressure of \mbar{<4e-10}. The preparation chamber is equipped with standard sputtering and e-beam annealing equipment, and has inlets for pure argon and nitrogen (99.999\%).

\figCoalescence

\Sssec{Note on the reported temperature}
We monitor the sample temperature during annealing by means of a pyrometer. Due to stray radiation originating from the filament behind the sample, the actual temperature readout, while reliable, is overestimated. In order to approximate the real temperature of the sample during annealing, we record the cooling curves after turning off the filament, and extrapolate back.

\figPercolation

We used a commercial \ce{Cu3Au} crystal grown by {\em Surface Preparation Laboratory}, which was cut along the (100) plane with $\sim$\angdeg{0.1} accuracy and polished to a roughness \SI{<0.03}{\micro\meter}.
Prior to growing the nitride islands, the crystal was cleaned with multiple rounds of argon sputtering at \kV{1} followed by annealing. This process was repeated until a clean surface with large plateaus was observed in STM images.

\figSpecs
\figManipulation

Nitrogen was subsequently implanted into the superficial layer by sputtering \ce{N+} ions onto the surface. We used an sputtering voltage of \V{500} and a current of \uA{1} to achieve coverages in the order of a monolayer per minute. To favor the mobility of the implanted nitrogen atoms and repair possible damage to the surface, we follow the sputtering by an annealing process, leading to the formation of a c(2x2)N reconstruction on the \ce{Cu3Au(100)} surface, similar to that reported for \ce{Cu(100)} \cite{leibsle_stm_1994}.

A \ce{Cu3Au} crystal can be in two distinct phases: an ordered \Lonetwo phase \cite{lu_electronic_1992, oguma_domain_2006} upon annealing below a critical temperature $T_c = \K{663}$ \cite{buck_order-disorder_1983, sundaram_order-disorder_1974, mannori_surface_1999}, and a disordered phase above this temperature \cite{morris_ordering_1974, jamison_polarized_1985}. While both phases have the FCC crystal structure, in the \Lonetwo phase the Au atoms are periodically distributed over the crystal whereas in the disordered phase they are not. The transition between the two phases is reversible \cite{keating_longrange_1951, benisek_vibrational_2015}: the crystal can be brought back into the \Lonetwo phase in a matter of hours by annealing at temperature near $T_c$ \cite{keating_longrange_1951, katano_lattice_1988}. The lattice constant of the disordered phase is slightly larger than the ordered \Lonetwo phase (\nm{0.3762} and \nm{0.3754}) respectively \cite{moffat_electrochemical_1991}).

\Sec{Results and discussions}
\label{sec:results}

In a first series of experiments, a clean and ordered \Lonetwo sample was sputtered with nitrogen for \seconds{45} at a current of \uA{0.8} and an accelerating voltage of \kV{0.5}. The sample was then annealed for \minutes{5}. The annealing temperature was kept at $T>T_c$ for only short periods of time, preserving the order in the bulk of the crystal. The surface is faster to both order and disorder when crossing the critical temperature, taking place on a broader temperature range \cite{sundaram_order-disorder_1973, houssiau_order-disorder_1996}.

\autoref{fig:ordered} shows the effect of different annealing temperatures (as determined via the process described above) on the size and distribution of the islands. The resulting islands vary in size from \nm{10} to \nm{100} in their longest direction where the largest islands appear only at a higher temperature. We observe a trend towards larger islands for higher temperatures. The edges of the island are mostly straight and oriented along the crystallographic axes (rotated between \angdeg{5} and \angdeg{10} clockwise relative to the image frame), as is observed on \ce{Cu(100)}. The island size is strongly increased with respect to the case of \ce{Cu(100)} owing to a reduced strain accumulation, due to the better match in lattice parameters.

A second series of experiments was performed after a prolonged high temperate treatment of the crystal. We annealed the crystal for 15 hours at \K{>900}. This temperature is well above the critical temperature of \K{663}, driving the crystal from the ordered \Lonetwo into a disordered FCC phase.

The disordered crystal was then prepared in a similar fashion as the ordered crystal. The amount of sputtered nitrogen is similar to the amount sputtered in the preparations in \autoref{fig:ordered} and the annealing time was kept unchanged at \minutes{5} for each preparation. The resulting surfaces at different annealing temperatures can be seen in \autoref{fig:disordered}.

The island sizes follow the same trend as on the ordered crystal, with islands ranging from \nm{10} to \nm{100} where larger islands are observed for higher temperatures. A major difference is found in the island geometry. In the ordered case the islands have mostly straight edges. In contrast, the islands of the disordered crystal are more rounded, showing no clear preferable orientation with regards to the crystallographic directions, indicating a more isotropic diffusion. This can be explained by disorder creating slight local variations in the lattice parameters and allowing the strain of the nitrogen reconstruction to be released. This strain release process could allow the reconstruction of islands without fundamental limits in their sizes.

Scanning the surface at higher bias voltage $V_b$ reveals features that were not evident for $V_b < \V{0.5}$. \autoref{fig:coalescence} shows the same island scanned at different bias voltages. At $V_b = \V{1.5}$ we observe bright spots appearing in the nitrogen reconstruction. At $V_b = \mV{10}$ and with atomic resolution, we can see that those bright spots corresponding to defects in the nitride lattice (see insets of \autoref{fig:coalescence}). The exact nature those defects is unknown, but they were observed only in the disordered phase. We suggest that they are \ce{Au} atoms that are incorporated into the copper-nitride layer as substitutions of \ce{Cu} atoms. In the \Lonetwo phase, every other layer in the (100) direction consists exclusively of \ce{Cu} atoms; after saturation with nitrogen, the surface is terminated on a \ce{Cu}-only layer \cite{niehus_surface_1993}.

The defect distribution gives us an indication on the formation and merging of islands. On round islands, the defects are mostly around the edges, where during growth new nitrogen joins the island and where the c$(2\times2)$ reconstruction is therefore not completed. Due to Brownian movement, the islands diffuse on the surface and will eventually collide. This process of coalescence is visible in \autoref{fig:coalescence}, where two islands were frozen in the process of merging. Longer annealing time or higher temperature would allow the island to properly merge and adopt a round shape. This establishes a clear relation between the elongation of the islands and the coalescence between multiple islands.

Raising the annealing temperature allows for faster dynamics, accelerating the island merging process which consequently leads to larger and rounder islands. \autoref{fig:percolation}a shows a \SI{\sim 145000}{\nano\metre^2} island observed on the disordered crystal. The total area of this island is an improvement of three orders of magnitude respect to the maximum area of a nitrogen islands on \ce{Cu(100)} \cite{oberg_control_2014-1}. However, in the inset of \autoref{fig:percolation}a, taken at higher bias, we can observe a high density of defects on the nitrogen reconstruction evenly distributed along the island, suggesting a higher Au-Cu substitution at the surface at elevated growth temperatures. Nonetheless, we successfully evaporated and manipulated \ce{Fe} atoms and on this island (white dots), and were able to engineer well-behaved spin structures (see \autoref{fig:manipulation}).

The area surrounding this island presents an irregular topography, highlighted in \autoref{fig:percolation}c-e, which we will denote as percolation regions. We observe this kind of behaviour for the preparations we performed at highest temperatures. They consist of a square pattern broken up by irregular channels connecting larger islands, as well as clean patches of exposed \ce{Cu3Au} surface. We have seen the percolation region to be unstable for $V_b > \V{4}$ both while scanning (see \autoref{fig:percolation}e, f) and during spectroscopy (see \autoref{fig:specs}).

We note that on the same sample preparation, it is possible to observe areas in the percolation regime and areas with regular nitride islands by macroscopically displacing the STM tip accros the sample surface, indicating that various phases can coexist on a single crystal. By starting the annealing at a higher temperature and gradually lowering the temperature, we are able to create round large islands with smooth nitrogen reconstruction, where the defects are mostly on the edges.

\autoref{fig:specs} shows constant current ${\rm d}I/{\rm d}V$ spectroscopy measurements on both regular and percolated areas. As seen in \autoref{fig:specs}b, the nitride islands in the regular region behave analogously to those reported for \ce{Cu(100)} \cite{choi_tunneling_2009}. In the percolated region (\autoref{fig:specs}c,d), three distinct phases are observed: two that behave similarly to the regular region (red, green) and the phase with the square pattern (black), where the spectroscopy is mostly featureless (apart of its instability).

The nitrogen islands on \ce{Cu3Au(100)} are suitable for adatom manipulation. We assembled many structures of \ce{Fe} adatoms --- from dimers to longer chains and blocks. Examples of such successfully assembled structure can be seen in \autoref{fig:manipulation}a. In \autoref{fig:manipulation}b we show spectroscopy measurements on the three atoms of a \ce{Fe} trimer, which are quantitatively the same as for a trimer assembled on nitride on \ce{Cu(100)} \cite{yan_control_2015}. Atomic manipulation is performed vertically, by moving an atom from the surface to the tip and subsequently form the tip to the surface on the desired position \cite{hirjibehedin_spin_2006}.

\Sec{Conclusion}
We have studied the growth of c($2\times2$) nitride islands on the \ce{Cu3Au}(100) crystal surface, which results in island sizes that are much larger than on the well-studied \ce{Cu}(100) surface. When the crystal is prepared in the ordered phase, we observe mostly rectangular nitride islands, which increase in size with temperature. On the disordered phase we see a similar relation between annealing temperature and island size, but in this case the islands are round, indicating that effects of strain due to lattice mismatch have diminished. Measurements at higher voltages reveal defects, the distribution of which gives information about the coalescence of islands during growth. The nitride islands on \ce{Cu3Au}(100) are found to be equally suitable for vertical manipulation of magnetic adatoms as their counterparts on \ce{Cu}(100).

\Sec{Acknowledgements}
We thank P. N. First for discussions and R. Toskovic, as well as A. M. Koning and J. W. van Dam for preliminary studies. This work was supported by the Netherlands Organisation for Scientific Research (NWO) and by the European Research Council.




\Sec{References}
\bibliographystyle{elsarticle-num}
\bibliography{Cu3Au-Prep-Paper}







\end{document}

%% file: figures.tex
\newcommand{\fullpageimg}{160mm}
\newcommand{\halfpageimg}{68mm}

\newcommand{\pA}[1]{\SI{#1}{\pico\ampere}}

\newcommand{\figOrdered}{
\begin{figure*}[h]
    \centering\includegraphics[page=1,width=\fullpageimg]{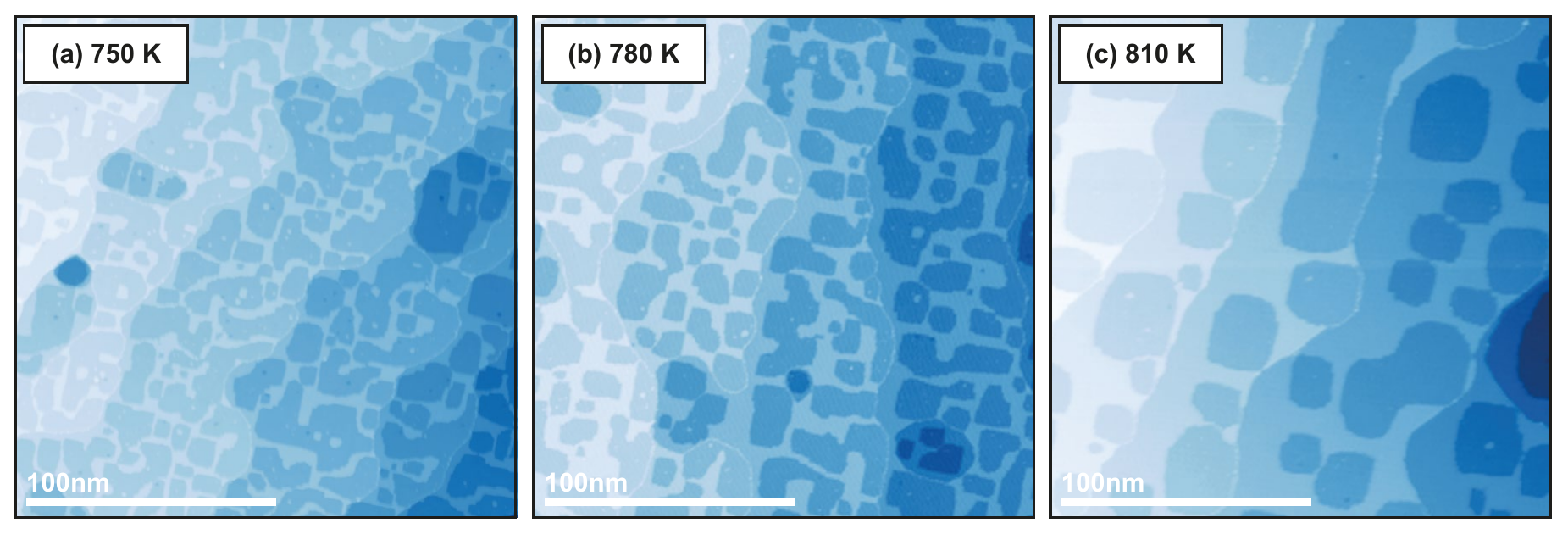}
    \caption{STM images of nitrogen islands on ordered \ce{Cu3Au(100)} after \minutes{5} of annealing at (a) \K{750} (b) \K{780} and (c) \K{810}. The images were acquire at a temperature of \K{1.5} with constant current at (a) \nA{0.1} and \V{0.2}, (b) \nA{0,1} and \V{1} and (c) \nA{0.4} and \V{0.2}. The nitrogen islands appear as darker areas surrounded by brighter areas, which are bare \ce{Cu3Au(100).}}
    \label{fig:ordered}
\end{figure*}
}

\newcommand{\figDisordered}{
\begin{figure*}[h]
    \centering\includegraphics[page=2,width=\fullpageimg]{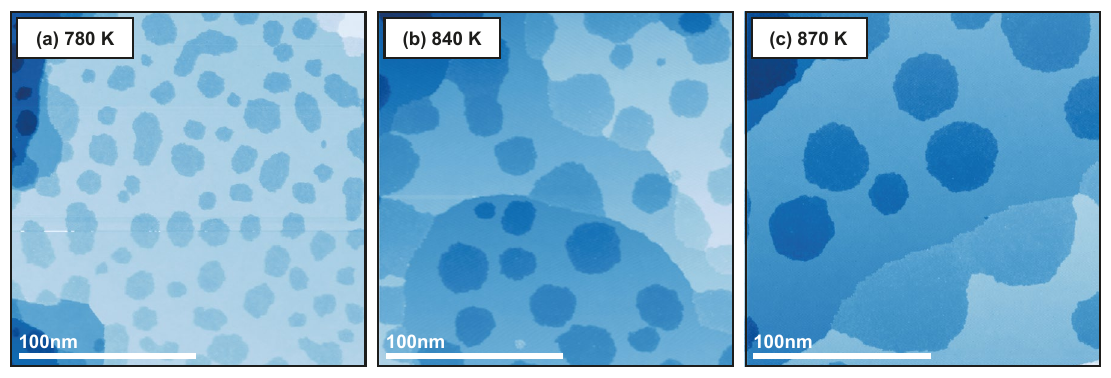}
    \caption{STM images of nitrogen islands on disordered \ce{Cu3Au(100)} after \minutes{5} of annealing at (a) \K{780} (b) \K{840} and (c) \K{870}. The images were acquire at a temperature of \K{1.5} with constant current at (a) \nA{0.1} and \mV{50}, (b) \nA{0.2} and \mV{100} and (c) \nA{0.2} and \mV{50}.}
    \label{fig:disordered}
\end{figure*}
}

\newcommand{\figCoalescence}{
\begin{figure*}[h]
    \centering\includegraphics[page=3,width=\fullpageimg]{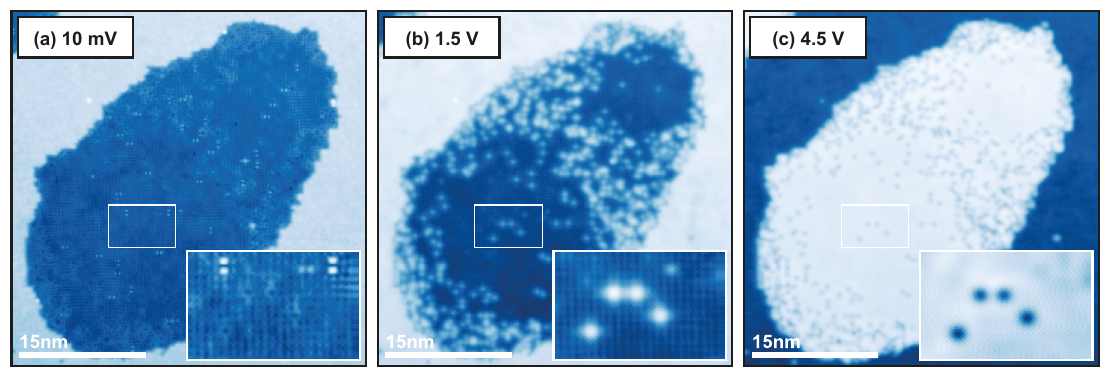}
    \caption{STM images of a nitrogen islands on disordered \ce{Cu3Au(100)}, scanned at a constant current of \nA{0.5} and a bias of (a) \mV{10}, (b) \V{1.5} and (c) \V{4.5}. The surface was annealed \minutes{5} at \K{840} after nitrogen sputtering. The insets are higher resolution scans of a section of the nitrogen island taken at identical measurement parameters. The corresponding area on the island is indicated by a white rectangle. For (a) and (b) the nitrogen islands are darker than the surrounding bare \ce{Cu3Au(100)}. This contrast is inverted for (c).}
    \label{fig:coalescence}
\end{figure*}
}

\newcommand{\figPercolation}{
\begin{figure*}[h]
    \centering\includegraphics[page=4,width=\fullpageimg]{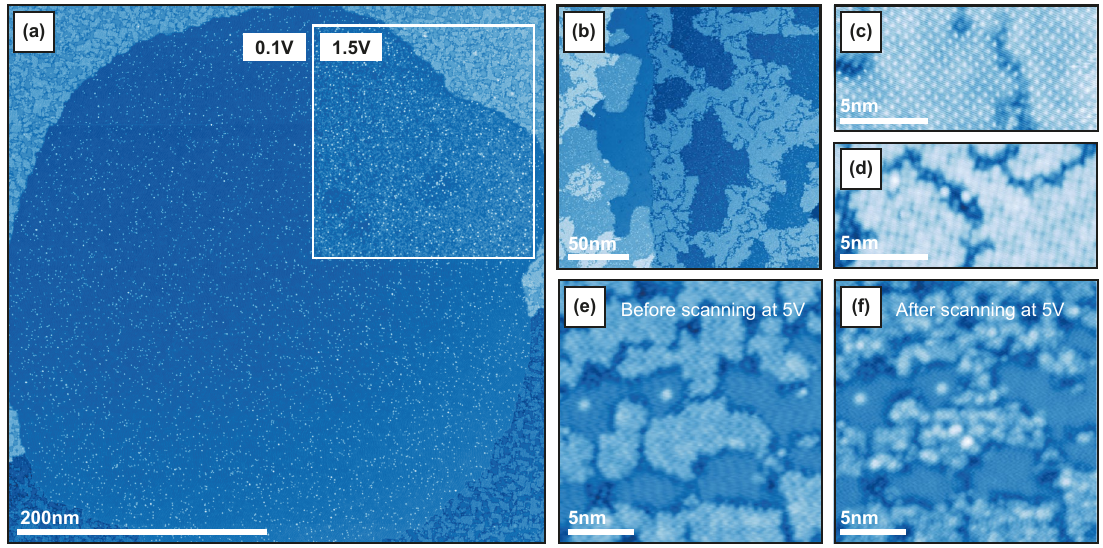}
    \caption{(a) Nitrogen island with an area of over \SI{145000}{\nano\metre^2} with \ce{Fe} adatoms evaporated onto it. (b-f) STM images of percolated areas. The sample was annealed for (a,e,f) \minutes{15} in three steps at \K{810}, \K{780} and \K{750}, (b) \minutes{20} at \K{930} and (c,d) \minutes{20} at \K{870}. The STM images were taken at (a) \nA{0.2} and \mV{100} (inset at \V{1.5}), (b) \nA{0.1} and \mV{200}, (c) \nA{0.1} and \mV{-20}, (d) \nA{0.2} and \mV{20} and (e, f) \nA{0.1} and \mV{100}. (e) and (f) show the same area before and after a scan of this area at \nA{0.1} and \V{5}.
    }
    \label{fig:percolation}
\end{figure*}
}

\newcommand{\figSpecs}{
\begin{figure}[h]
    \centering\includegraphics[page=5,width=\halfpageimg]{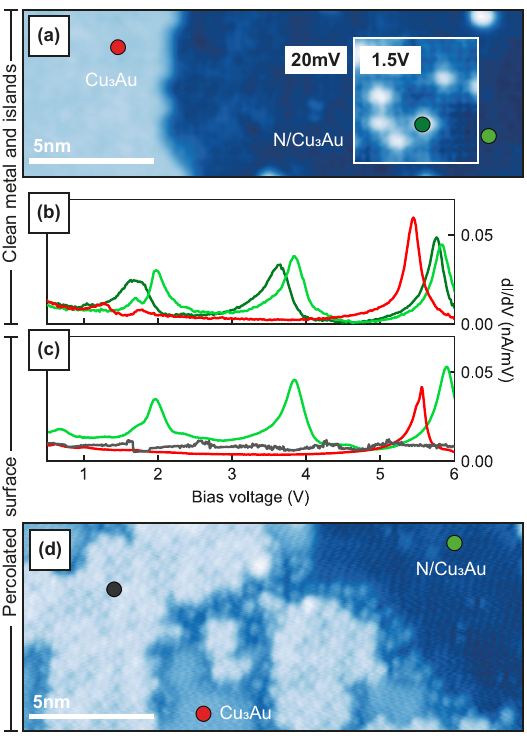}
    \caption{Tunneling specstroscopy measurements on regular and percolated areas. (a) STM image of a regular nitrogen island on disordered \ce{Cu3Au}. (b,c) ${\rm d}I/{\rm d}V$ spectroscopy measurements at constant current, taken at different positions on regular (b) and percolated (c) areas. Measurement locations are indicated with corresponding colours in (a) and (d). (d) STM image taken in a percolated area. STM scans were taken at \K{1.5} and (a) \pA{100} and \mV{20} (inset at \V{1.2}) and (d) \pA{200} and \mV{100}.
    }
    \label{fig:specs}
\end{figure}
}

\newcommand{\figManipulation}{
\begin{figure}[h]
    \centering\includegraphics[page=6,width=\halfpageimg]{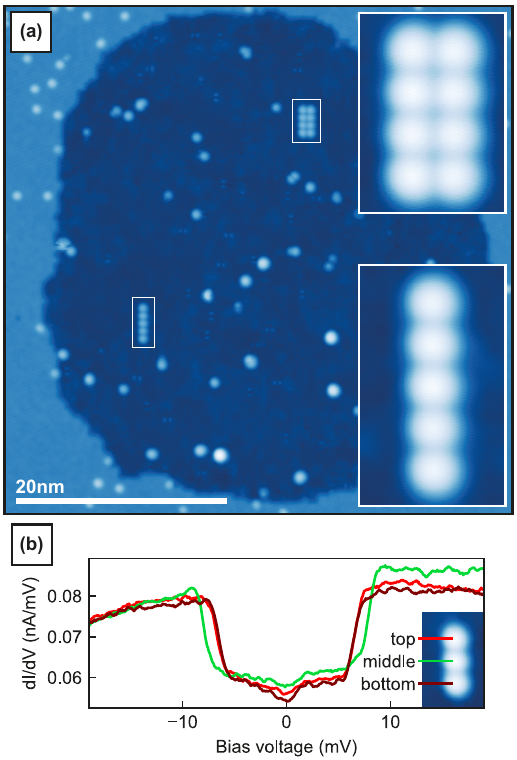}
    \caption{(a) $1\times5$ and $2\times4$ structures of \ce{Fe} atoms similar to \cite{loth_bistability_2012} assembled through vertical manipulation on a \SI{\sim 5000}{\nano\metre^2} nitrogen island on \ce{Cu3Au}. (b)  ${\rm d}I{\rm d}V$ spectroscopy measurements taken on each of the atoms of a $1\times3$ Fe structure built on an island of similar size as in (a). Spectroscopy was taken in the center of each atom with initial settings of \pA{800} and \mV{-20}.  All structures were built on the same sample, which was annealed at \K{810}, \K{780} and \K{750} for \minutes{5} at each temperature. The STM images were taken at (a) \pA{50} and \mV{20} and (b) \pA{200} and \mV{50}.
    }
    \label{fig:manipulation}
\end{figure}
}